\documentclass[aps,pra,preprint,showpacs,preprintnumbers,amsmath,amssymb]{revtex4-1}
\usepackage{amsmath}
\usepackage{latexsym}
\usepackage{txfonts}
\usepackage{graphicx}
\usepackage{amssymb}
\usepackage{float}
\usepackage{subcaption}

\begin{document}

\title{Orbital angular momentum of a $\pi$-pulse emission by dense relativistic cold electron beam }
\author{J. A. Arteaga$^1$, A. Serbeto$^1$, J. T. Mendon\c{c}a$^2$, K. H. Tsui$^1$, L. F. Monteiro$^1$ }
\affiliation{$^1$Instituto de F\'isica, Universidade Federal Fluminense, Campus da Praia Vermelha, Niter\'oi, RJ,  24210-346, Brasil,}
\affiliation{$^2$IPFN, Instituto Superior T\'ecnico, Universidade de Lisboa, 1049-001 Lisboa, Portugal}
\email{johny@if.uff.br}

\begin{abstract}

Using a quantum fluid model, a set of three paraxial coupled equations to describe the FEL instability is derived. These equations  are solved numerically considering Laguerre-Gaussian modes as the initial conditions to study the transfer of orbital angular momentum (OAM) between them, which satisfies the total angular momentum conservation as matching condition. The amplification and compression of the output radiation is observed and using the separation of variables method, the set of coupled equations, which describes a $\pi$-pulse solution, is obtained in such a way that a transverse overlapping factor, R, governs the energy exchange efficiency of the process.
\end{abstract}

\maketitle

\section{Introduction}

The search for ultra high intensity and short wavelength  emission produced by a Free-electron Laser  (FEL), which was conceived by Madey and co-workers \cite{madey1971},   has taken a considerable research effort in order to reduce its cost and its size.  This endeavour has suggest us to built a FEL scheme in such a way that it can be used as a compact system. In order to reach this objective, many authors have suggested the use of an intense laser field as the wiggler in substitution to the static magnetic wiggler array. In the optical wiggler configuration, a counter propagating laser field interacts with a relativistic electron beam through the stimulated scattering mechanism such that an induced stimulated intense radiation is emitted by the FEL system \cite{sprangle2009, Andriyash}.

 Although the classical description for the FEL dynamics has been used to describe completely the system, if the photon momentum recoil is not greater than the beam momentum spread, this classical model fails in the case of emission of very short wavelength, for instances, X-Ray and $\gamma$-Ray emissions, since in this situation the electron and photon momentum recoils could be of the same
order of magnitude. Theoretically, in this case, a quantum mechanical description should be used in order to describe correctly the FEL instability when the effect of the photon recoil is considered. Inspired by this effect of very short wavelength on FEL dynamics, Bonifacio et. al.\cite{Boni2005} have proposed a quantum mechanical model such that the electrons of the relativistic beam are described by a macroscopic matter-wave function, which obeys the relativistically corrected Schrodinger-like equation, which was previously derived by Preparata\cite{preparata} using the quantum field theory approaches, where the solution describes the FEL instability. However, an alternative quantum mechanical approach has been suggested in Ref.\cite{serbetoqfel, lfmonteiro} to study the FEL instability. In this model a quasi-classical quantum hydrodynamical model is derived such that the relativistic electron beam is considered as a quantum plasma fluid interacting with electromagnetic waves through the three-waves process. Based on this set of basic equations derived in Refs.\cite{lfmonteiro, serbetoqfel, guarumo}, a stimulated coherent $\pi$-pulse output radiation is generated as a result of the interaction of the dense relativistic cold electron beam with a counter propagating optical wiggler. 

Recently, from the seminal idea of Allen \cite{allen}, several research efforts have been developed to understand and describe the properties of an electromagnetic wave, which carries orbital angular momentum (OAM) during its interaction with the matter. In this case, the light beam is well described by the Laguerre-Gaussian (LG) mode, which has a donut-like shaped intensity profile and a corkscrew-like phase. The common mechanism to produce OAM visible light is based on the spiral phase plates or computer-generated holograms. In Ref. \cite{titonature}, the authors  present a mechanism to transfer OAM to a high intensity output light through the stimulated Raman backscattering in a plasma. Using a helical relativistic electron beam, Hemsing et al. \cite{hemsingnature} produce a high brightness OAM X-ray emission in the SLAC FEL. At the FERMI FEL a intense  extreme-Ultraviolet vortex light beam was obtained  using a helical ondulator  and spiral zone plate \cite{felvortex}. Here,
using the three-wave interaction parametric process in a quantum plasma fluid model and making use of the slowly-varying envelope approximation (SVEA), we explore the capability to transfer OAM from an optical wiggler and scattered radiation to the electron beam density perturbation, and from optical wiggler and the perturbed relativistic electron beam to the emitted output radiation.

\section{Model}

It is well known that an electromagnetic wave can be described by a vector potential, which obeys the following  wave propagation equation,
\begin{equation}
\left( \nabla^2_\perp + \frac{\partial^2}{\partial z^2} + \frac{\partial^2}{\partial t^2}   \right) \vec{A}_\perp (\vec{r}_\perp, z,t) = \frac{4\pi e^2 n }{\gamma m_e c} \vec{A}_\perp(\vec{r}_\perp,z,t), 
\end{equation}
where $\vec{A}_\perp = \vec{A}_w + \vec{A}_s  = \hat{e} \{A_w \,\textnormal{exp}[i(-k_w z + \omega_w t )] + i A_s\, \textnormal{exp} [i(k_s z + \omega_s t )]  \}/\sqrt{2}$ is the total circularly polarized  electromagnetic wave propagating in the $\hat{z}$ direction due to optical  wiggler and a  counter propagating  scattered radiation.
Considering the electron beam density perturbation as $\delta n /n_b = n\,\textnormal{exp}(i(-k_l z- \omega_l t))$ and neglecting the non resonant terms on the FEL matching conditions, and using  the slow-varying envelope approximation ($\partial^2_z \ll k_{s,w} \partial_z$, $\partial^2_t \ll \omega_{s,w} \partial_t  $),  we can obtain the evolution equation for each normalized electromagnetic mode 
\begin{eqnarray}
\left( i \frac{c^2}{2\omega_s} \nabla^2_\perp - \frac{c^2 k_s}{\omega_s} \frac{\partial}{\partial z} + \frac{\partial}{\partial t} \right) a_s &=& -\frac{\omega_p^2}{2 \omega_s \gamma_e}\, n^*\, a_w,  \\ 
\left( i \frac{c^2}{2\omega_w} \nabla^2_\perp + \frac{c^2 k_w}{\omega_w} \frac{\partial}{\partial z} + \frac{\partial}{\partial t} \right) a_w &=& \frac{\omega_p^2}{2 \omega_w \gamma_e}\, n\, a_s,
\end{eqnarray}
where the dispersion relation of each electromagnetic mode,  $ \omega^{2}_{s,w} = \omega_p^2/\gamma_e + c^2 k^2_{s,w} $, has been used.

Making use of the basic equations derived in Refs. \cite{lfmonteiro, serbetoqfel}, the evolution equation for the electron beam density perturbation, $\delta n/n_{b}$, is given by
\begin{equation}
\frac{\partial^2}{\partial t'^2}\frac{\delta n}{n_b} + \frac{\hbar^2}{4 m_e^2 \gamma_e^4 } \nabla^2 \left( \frac{1}{\gamma_e^2}\frac{\partial^2}{\partial \zeta^2} + \nabla_\perp^2 \right) \frac{\delta n}{n_b} + \frac{\omega_p^2}{\gamma_e^3} \frac{\delta n}{n_b} = i \frac{c^2}{2\gamma_e^4} \nabla^2(a_w \, a_s^*\, e^{i \theta_n}), 
\end{equation}
where $\nabla^2 = \partial_\zeta^2 + \nabla_\perp^2$ and $\zeta = z - v_{e}t$. Considering $\delta n /n_b = n\,\textnormal{exp}(i(-k_l \zeta- \Omega_l t'))$ and assuming the slowly-varying envelope approximation (SVEA) in the longitudinal direction and in time, such as $k_l^3 \partial_\zeta \gg k_l^2 \partial^2_\zeta \gg k_l \partial^3_\zeta \gg \partial^4_\zeta $  and $k_l^2 \nabla^2_\perp \gg k_l \partial_\zeta \nabla^2_\perp \gg  \partial^2_\zeta \nabla^2_\perp \gg \nabla^4_\perp $, the evolution equation for $\delta n/n_{b}$ can be rewritten in the following form,
\begin{equation}
\left[ S_Q^2 \left( -i \frac{1}{2\Omega_l} \nabla_\perp^2 - \frac{2}{\gamma_e^2}\frac{k_l}{\omega_l} \frac{\partial}{\partial \zeta} \right) + \frac{\partial}{\partial t'} \right] n = \frac{k_l^2 c^2}{4 \Omega_l \gamma_e^4} \, a_w a_s^*
\end{equation}
where $S_Q^2 \equiv \hbar^2 k_l^2/2  m^2 \gamma^4$ is the quantum plasma wave velocity and $\Omega_l^2 = (\omega_l - k_l v_e)^2 = \omega_p^2/ \gamma^3 + c^2 \lambdabar_c^2 k_l^4/ 4\gamma_e^6$ is the Doppler-shifted quantum plasma dispersion relation, which have been used in order to derive the above density evolution equation. To express  all the envelope equations as function  of the scattered radiation group velocity, we will make use of the following space and time  transformation: $ \bar{z} = z + (c^2k_s/\omega_s) t$ and $\bar{t} = t'$,  for the electromagnetic wave equations, and $\bar{z} = \zeta + (c^2k_s/\omega_s - v_e) t'$ and $\bar{t} = t' $, for the perturbed electron density equation. Therefore, the coupled set of non linear equations, which describe the  three-wave interaction process, reads as 
\begin{eqnarray}
\left( i R_s \nabla^2_\perp + \frac{\partial}{\partial \bar{t}} \right)  a_s(r_\perp,\bar{z},\bar{t})  &=&  -\frac{\omega_p}{2 \gamma_e \omega_s}\, n^*\, a_w  = - c_s \, n^*\, a_w \label{as}, \\ 
\left[  \left( i R_w\nabla^2_\perp +  \frac{\partial}{\partial \bar{t}}\right)  + (\beta_s + \beta_w)\frac{\partial}{\partial \bar{z}}  \right] a_w(r_\perp,\bar{z},\bar{t}) &=&  \frac{\omega_p}{2 \gamma_e \omega_w}\, n\, a_s  = c_w \, n\, a_s \label{aw}, \\
\left[ \left(-i R_n   \nabla_\perp^2 + \frac{\partial}{\partial \bar{t}} \right) -\beta_Q \frac{\partial}{\partial \bar{z}} + (\beta_s - \beta_e)\frac{\partial}{\partial \bar{z}} \right] n(r_\perp,\bar{z},\bar{t}) &=& - \frac{c^2 k_l^2}{4 \gamma_e^4 \omega_p |\Omega_l| }\,a_w\, a_s^* = -c_n \,a_w\, a_s^*  \label{n},
\end{eqnarray}  
where $R_{s,w} = \omega_p/2 \omega_{s,w}$, $R_n = S_Q^2 \omega_p/(2c^2\Omega_l)$, $\beta_Q = 2 S_Q^2  k_l/(c \gamma_e^2 \Omega_l) $, and the space and time variables are re-scaled with $\omega_p/c$ and $\omega_p$ respectively. 
\\
\\
It is well known that the solution of the  cylindrical uncoupled  paraxial equation, given by $(i R_{s,w} \nabla^2_\perp + \partial_t)a_{s,w} = 0 $,  can be expanded in a  family of orthogonal LG modes, with  each of them  distinguished by the subscripts $\{p,\ell\}$, where $\{p\}$ stands for the number of rings in the intensity profile, and $\{\ell\}$ represents the amount of OAM carried by the mode. Thus, each envelope which is prepared in a LG mode has the following form
\begin{equation}
 a\, (r, \varphi,t)^{|\ell|}_{p}= A_0(z)\, w_{0s,w} \, \textnormal{LG}^{|\ell|}_{p}(r,\varphi,t)  \, \,\textnormal{exp}\left(\frac{4iR_{s,w} t}{z_{0s,w}^2 + 16 R_{s,w}^2 t^2} r^2\right) \,\, \textnormal{exp} \left[ -i( 2p + |\ell| + 1)\arctan(t/z_{0s,w}) \right],    
\end{equation} 
with 
\begin{equation}
\textnormal{LG}^{|\ell|}_{p}(r,\varphi,t) = \sqrt{\frac{2p!}{\pi w^2_{s,w} (p + |\ell|)!}} \left(  \frac{\sqrt{2} r}{w_{w,s}} \right)^{|\ell|} L^{\ell}_p \left( \frac{2 r^2}{w^2_{w,s}} \right) \, e^{- \frac{r^2}{w^2_{w,s}}} \, e^{i \ell \varphi},
\end{equation}
where $w_{s,w}^2= z_{0s,w} (1 + 16 R^2_{s,w} t^2/z_{0s,w}^2) $ gives the temporal evolution of the transverse waist, and $z_{0s,w} = w_{0s,w}^2$ represents the normalized Rayleigh length of the scattered radiation and optical wiggler respectively. The orthogonality condition of the Laguerre-Gaussian functions remains as follows  $\int { d^2 r }\, \textnormal{LG}^{|\ell'|*}_{p'} \textnormal{LG}^{|\ell|}_{p}  = \delta_{\ell, \ell'} \delta_{p,p'}$.

\section{Numerical Results}

In order to solve numerically the set of Eqs.(\ref{as})-(\ref{n}), we will assume an initial optical wiggler with a transverse Gaussian shape ($\ell_w=0,p=0$) and a constant longitudinal profile. The initial scattered radiation will be assumed to have an OAM ($\ell_s = 1 $) and a longitudinal Gaussian shape, and a null density perturbation, namely, 
\begin{eqnarray}
a_w(r,\varphi ,z,t=0) &=& A_w \, e^{-r^2/w_0^2} \label{aw1},\\
a_s(r,\varphi, z,t=0) &=& A_s e^{(z-z_0)^2/\sigma_z^2}\, e^{-r^2/w_0^2} \,\frac{1}{\sqrt{\pi}} \frac{\sqrt{2} r}{w_0} \, e^{i  \varphi} \label{as1},\\
n(r,\varphi, z,t=0) &=& 0 \label{n1}.
\end{eqnarray}
Throughout this work, for simplicity, the longitudinal interaction size is considered  less than the  minimum Rayleigh length of each  mode to avoid  optical diffraction and also $\{p\} = 0$ in all situations.
\\ 

Fig.1 represents the initial configuration given by the set of  Eqs. (\ref{aw1})-(\ref{n1}). The initial scattered radiation  envelope, $|a_s| = \sqrt{|a_s|^2}$, is contained in the interior of the simulation box and the longitudinal cross-section of the constant optical wiggler is located in the interior face of the box. On the left face of the box, we represent the transverse cross-section of the maximum value of the seed $|a_s|$. It should be noticed from the picture that the maximum initial value of this seed is located at the longitudinal position $z_0 = 0.4$. We can observe a typical donut-like shaped of the LG modes when projected onto this face. The method to quantify the amount of OAM carried by the modes is by counting the number of symmetrical maximum of the squared real part envelope amplitude Re$\{a_s\}^2$ and Re$\{n\}^2$, which is represented on the right hand side of this figure. For the transverse cross-section of the squared real part, Re$\{a_s\}^2$, of the seed $a_{s}$, taken at  $z_0=0.4$, we can observe the existence of two symmetrically separated maxima, showing the scattered radiation with OAM $|\ell_s|=1$. Fig.2 gives us the spatial evolution of the solution for the set of Eqs.(6) - (8) in an advanced time of the interaction. In Figs.2(a) and 2(b), we observe the division of the optical wiggler and the scattered radiation into sub-pulses. The transverse shape of the modes is constant with a decreasing waist size during the process. Inside the box represented in Fig.2(a), the leading pulse of the scattered radiation has  increased approximately twenty times in amplitude, while the longitudinal and transverse waist have experienced a decrease during the interaction time. The maximum value of the transverse cross-section of the seed $|a_s|$ is represented in the lower face of Fig.2(a). It should be noted that the narrow shape is due to the strong compressional effect. The right side of the box in Fig.2(a) represents the longitudinal cross-section of the optical wiggler taken at the maximum value of scattered radiation. Here, we witness that the maximum values of the radiation pulses are located at the regions of minimum amplitude of the optical wiggler, therefore leading to the optical wiggler depletion profile, as given in Fig.2(b). Fig.2(c) represents the behavior of the corresponding beam density perturbation during the interaction.
Since in the three wave interaction process, the exchange of energy among an electromagnetic wave (pump wave) to a stimulated electromagnetic wave (scattered radiation) has to be mediated by a perturbed electron density oscillation, which in principle can start from a very small amplitude (noise) and evolve to a profile as shown in Fig.2(c).
\begin{figure}[h!]
\begin{subfigure}{1\textwidth}
  \centering
 \includegraphics[width=\linewidth ]{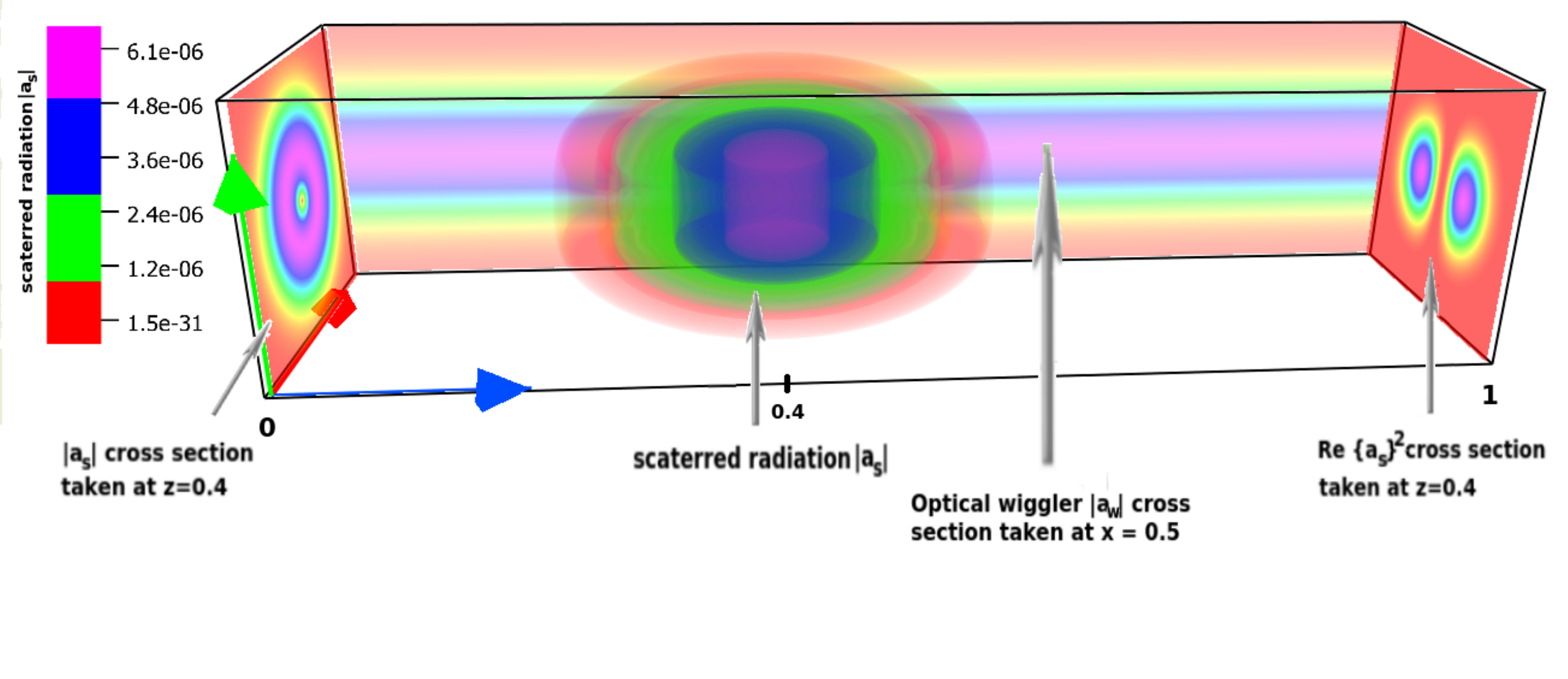}
\end{subfigure}%
\caption{Initial conditions for the optical wiggler and scattered radiation. The $x,y,z$ axes are represented by red, green, and blue arrows respectively. The Gaussian constant wiggler has an amplitude $A_w = 10^{-4}$, and the OAM scattered radiation has an initial amplitude $A_s = 10^{-5}$ located at $z_0 = 0.4$.}
\label{fig:nlinear}
\end{figure}
\begin{figure}[h!]

\begin{subfigure}{0.3\textwidth}
  \centering
  \includegraphics[width=\linewidth ]{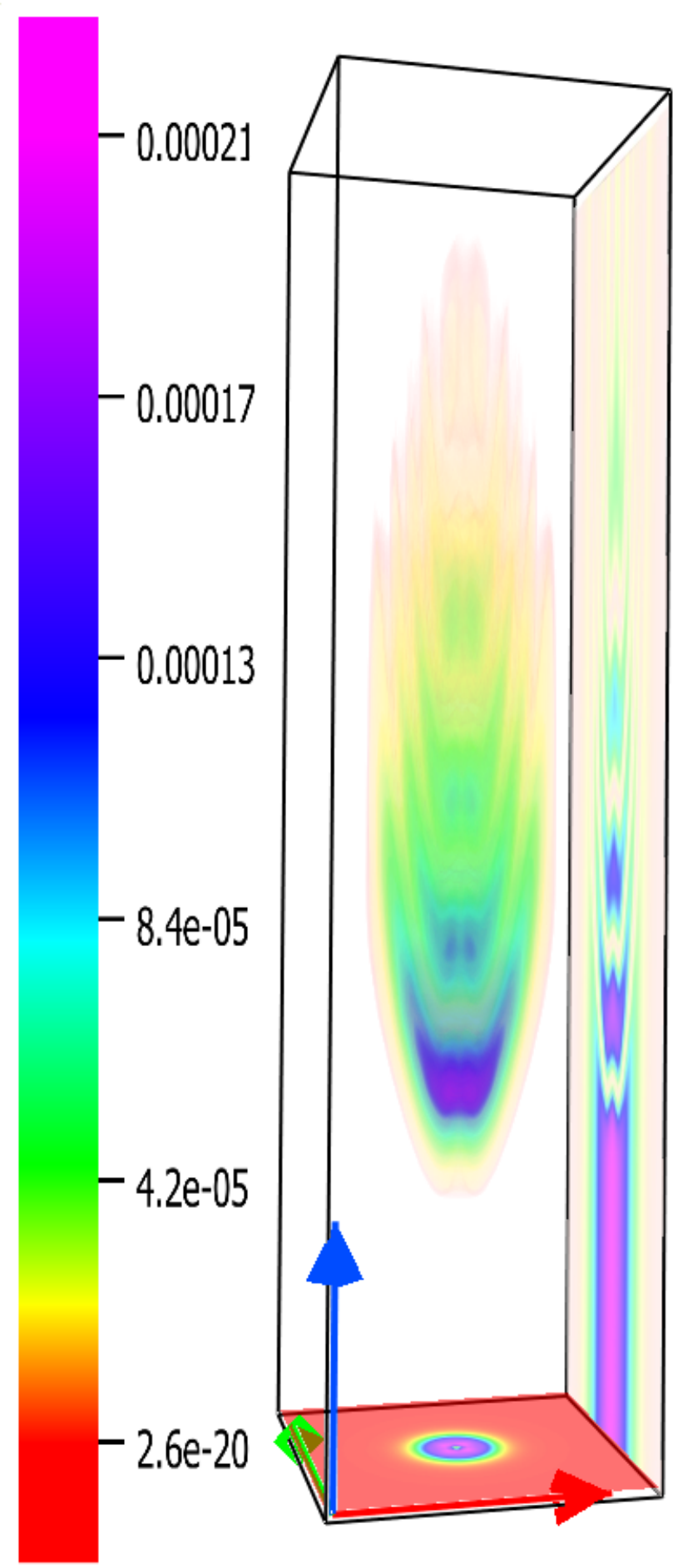}
 \caption{}
\end{subfigure}
\begin{subfigure}{0.3\textwidth}
  \centering
  \includegraphics[width=1.0 \linewidth]{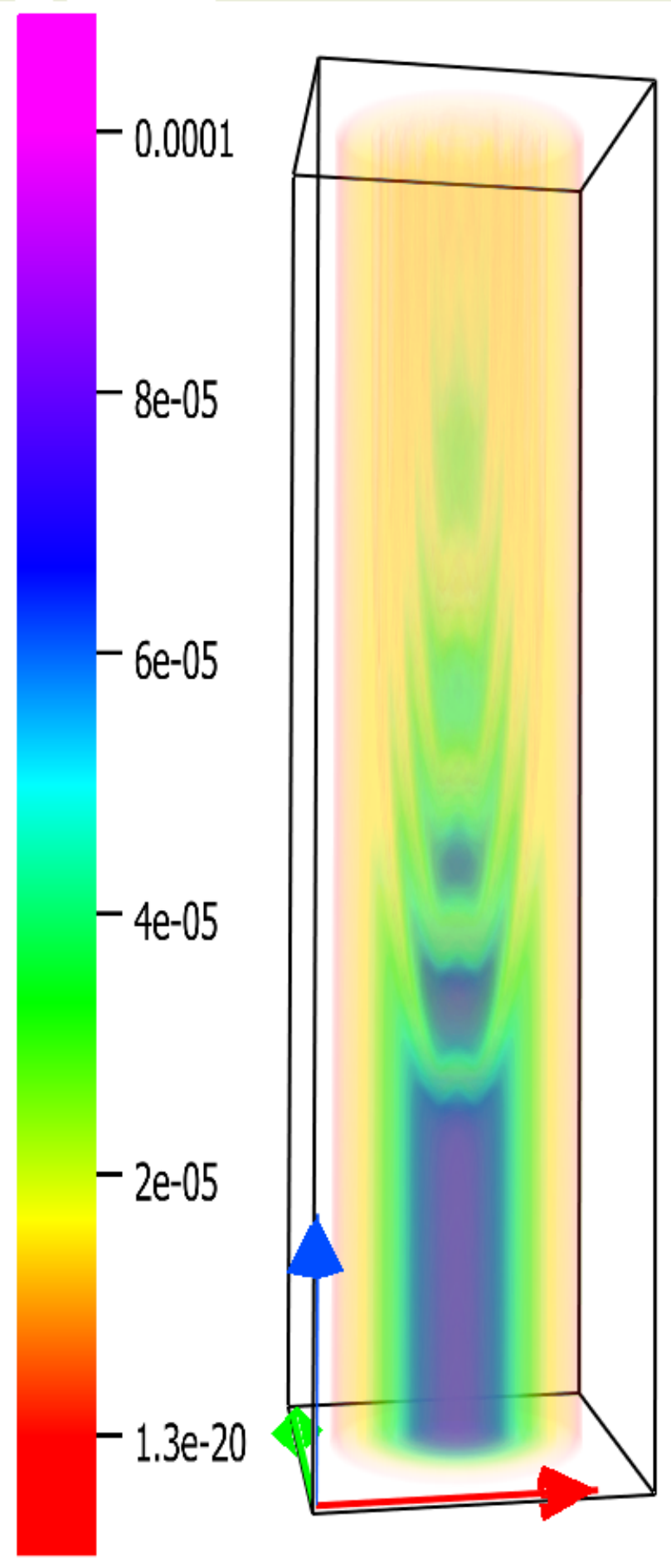}
 \caption{}
\end{subfigure}
\begin{subfigure}{0.27\textwidth}
  \centering
  \includegraphics[width=1.0 \linewidth]{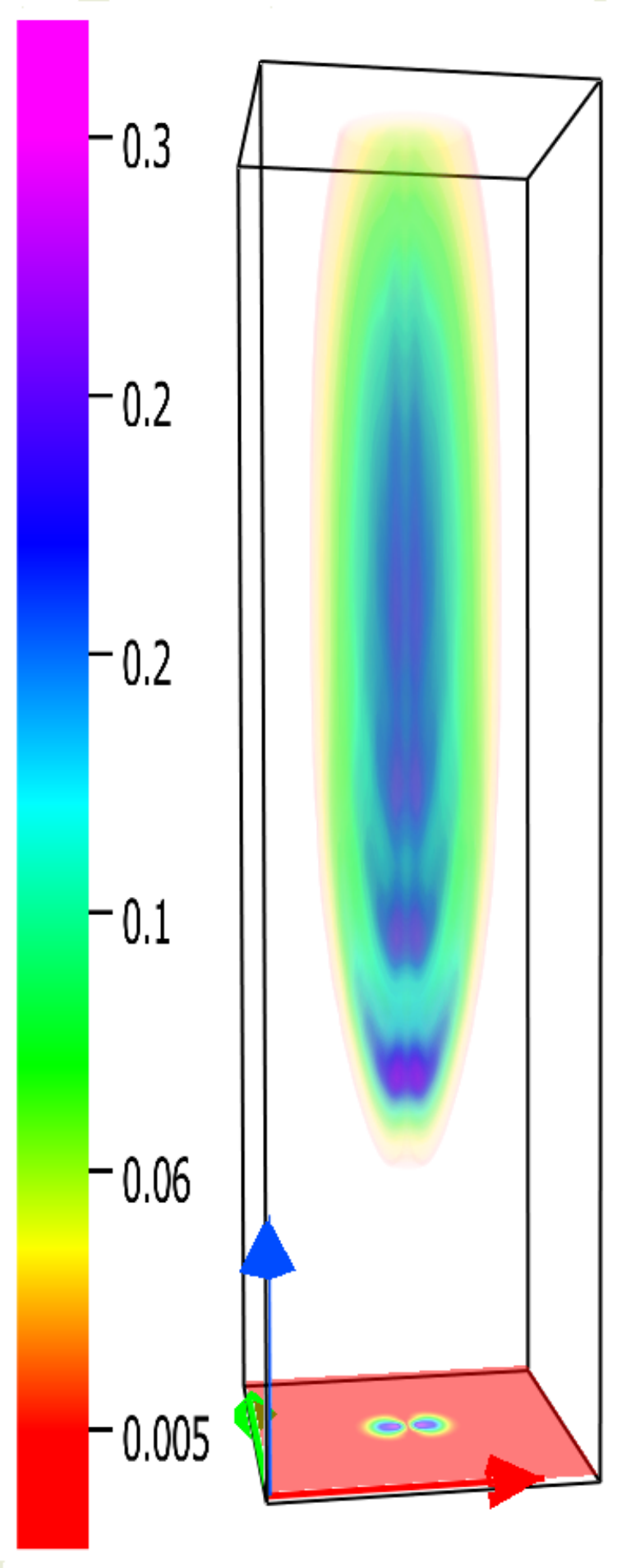}
 \caption{}
\end{subfigure}
\caption{Spatial evolution in one advanced time of interaction for the initial configuration described in Fig. 1: $\textbf{(a)}$ for the absolute value of the seed envelope, $|a_s|$, with the wiggler pump depletion represented on right face of the box, $\textbf{(b)}$ the evolution of the optical wiggler, $|a_w|$, $ \textbf{(c)}$ the evolution of the density perturbation, $|n|$, in the top of this figure given by $Re(n^2)$. The parameters used in the simulation are $n_b = 10^{20}cm^{-3}$, $\gamma_e = 5.0$, $\lambda_w = 550nm$.}
\label{fig:nlinear}
\end{figure}
Here, starting from the first maximum of the scattered radiation, we observe that the electron beam density is perturbed over the entire posterior region with a maximum amplitude at the same place of the seed $|a_s|$.            
According to these results, the process of energy exchange among the modes and the transverse compression of the leading pulse radiation have similarity to the longitudinal features of the $\pi$-pulse emission by a FEL in a three-wave process studied in \cite{guarumo}. In addition, the maximum value of the transverse cross-section of the squared real part of the density perturbation, Re$\{n\}^2$, is presented in the lower face of Fig.2(c), which shows us two symmetrically well separated narrow maxima, that indicate the presence of OAM ($|\ell_n| = 1$) on the beam mode, i.e, the electron beam twist. \\

To elucidate the characteristics obtained so far, we use the following variable separation: $a_s = T_s(r,\varphi,t) A_s(t,z)$, $a_w = T_s(r,\varphi,t) A_s(t,z)$, and $n = T_n (r,\varphi,t) N(t,z)$, where the $T(r,\varphi,t)$ functions satisfy the uncoupled paraxial equation for each envelope. Considering $w_{s,w,n} = w_0$ and neglecting diffraction of the modes ($t/z_0 \ll 1$), and taking into account the orthogonality condition of the LG modes, we get the following set of equations
\begin{eqnarray}
\partial_{\bar{t}} A_s &=& - c_s\, R\, A_w N^*, \nonumber \\
\left[\partial_{\bar{t}}  + (\beta_s + \beta_w)\partial_{\bar{z}}  \right] A_w &=& c_w\, R^*\, N A_s ,\nonumber \\
\left[\partial_{\bar{t}}  + (\beta_s - \beta_Q - \beta_e)\partial_{\bar{z}}  \right] N &=& c_n\, R \,A_w A_s^*.  \nonumber
\end{eqnarray}
Here, $R \equiv \int d^2 r\, \textnormal{LG}^{|\ell_s|*}_p\,\textnormal{LG}^{|\ell_w|}_p\, \textnormal{LG}^{|\ell_n|*}_p$    represents the overlapping integral of the transverse structure of the three interacting modes. It is clear from this set of equations with transverse coupling, the analogy of this process with the $\pi-$pulse description in Ref.\cite{guarumo}. Furthermore, looking for the azimuthal dependence of the coupling term $R$, we can derive the following angular momentum matching condition \cite{titoprl}
\begin{equation}
  \ell_w = \ell_n + \ell_s. \label{matching}
\end{equation}
\begin{figure}[h!]

\begin{subfigure}{0.3\textwidth}
  \centering
 0 \includegraphics[width=\linewidth ]{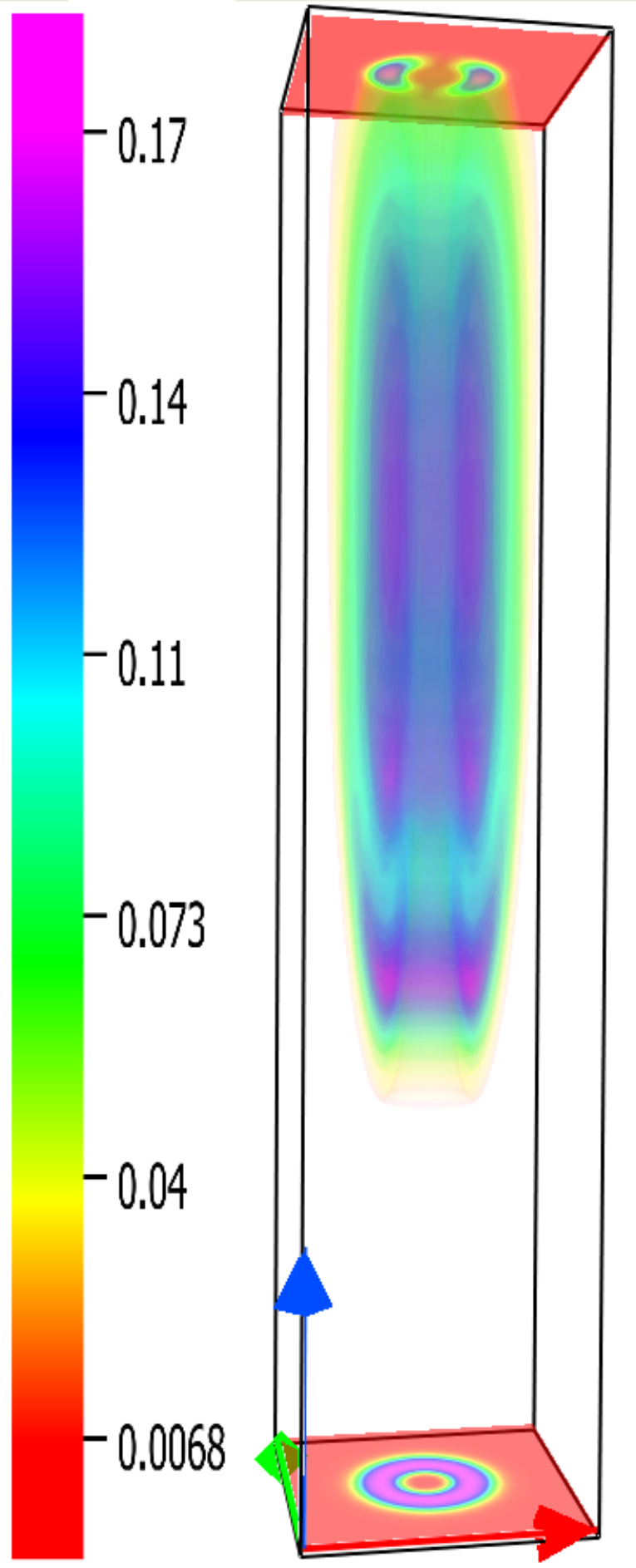}
 \caption{}
\end{subfigure}
\begin{subfigure}{0.3\textwidth}
  \centering
  \includegraphics[width=1.0 \linewidth]{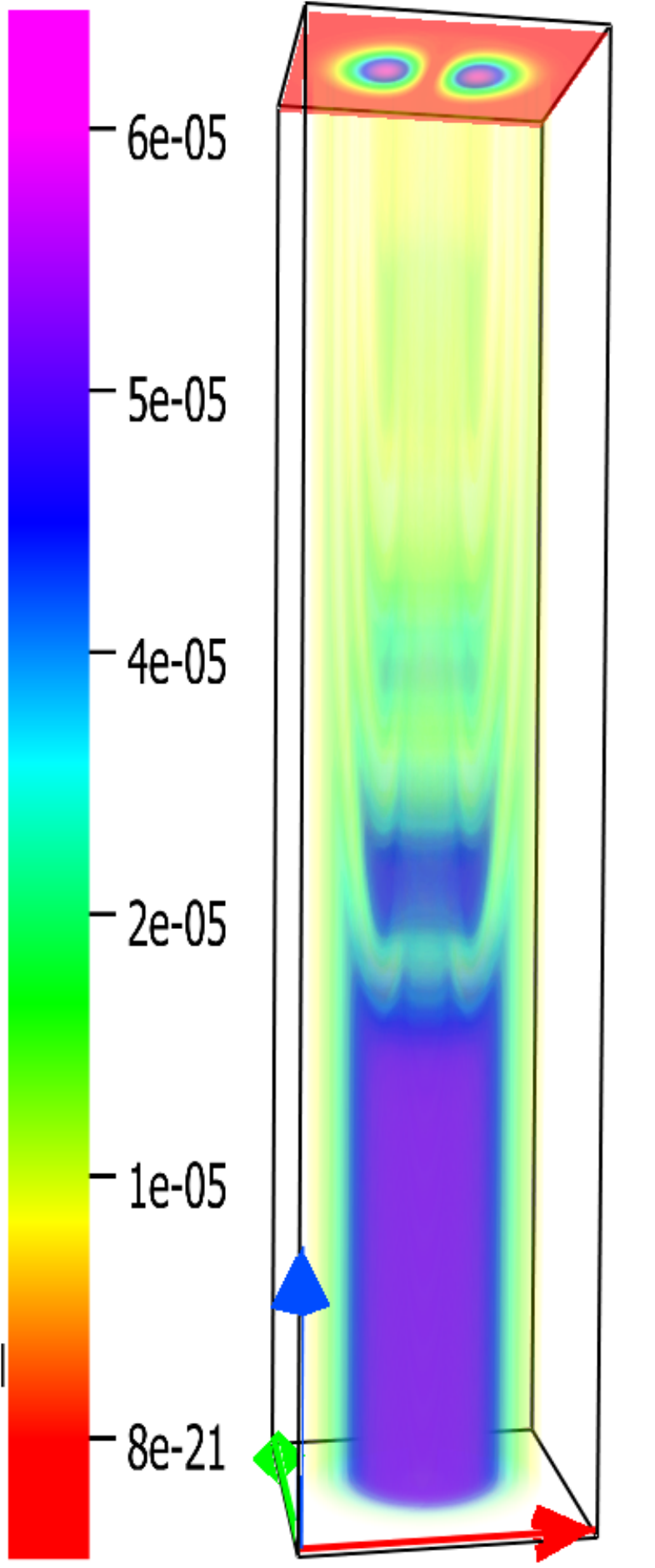}
 \caption{}
\end{subfigure}
\begin{subfigure}{0.27\textwidth}
  \centering
  \includegraphics[width=1.0 \linewidth]{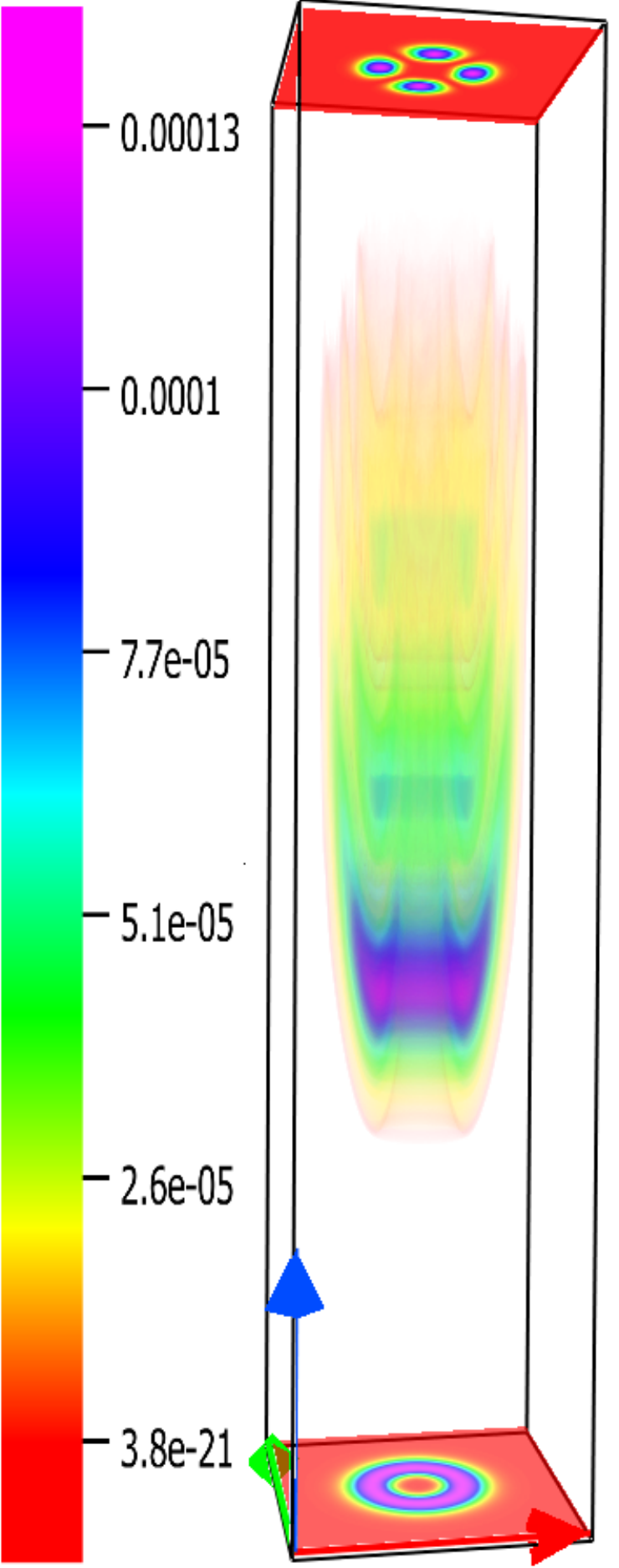}
 \caption{}
\end{subfigure}
\caption{Spatial evolution taken at the same time of interaction of figure 2 for the initial configuration described in Eqs.(\ref{aw2})-(\ref{n2}). \textbf{a)} The lower transverse face presents the transverse cross-section of $|n|$ taken at its longitudinal maximum. The top transverse face presents the transverse cross-section of Re$\{n\}^2$ taken also at the longitudinal maximum of $|n|$. \textbf{ b)} The top transverse face presents the transverse cross-section of Re$\{a_w\}^2$ taken at $z=0$. \textbf{c)} The lower transverse face presents the transverse cross-section of $|a_s|$ taken at its longitudinal maximum. The top transverse face presents the transverse cross-section of Re$\{a_s\}^2$ taken at the longitudinal maximum of $|a_s|$. The FEL parameters used here are the same as the case presented in the figure 2.}
\label{fig:nlinear}
\end{figure}
Since both scattered radiation and perturbed beam density are described by a first order perturbation theory of the optical wiggler, we can think of a scenario with a non null perturbed density amplitude and null scattered radiation amplitude in order to produce the same features of $\pi-$pulse solution coming from the usual configuration as given by Eqs.(\ref{as1})-(\ref{n1}). This alternative scenario was implemented and corroborated in \cite{plasmaseed} using an initial non-zero perturbed Langmuir density in a one dimensional backward Raman amplifier. Inspired by this second situation with an initial constant optical wiggler carring an OAM $\ell_w = 1$ and a finite perturbed beam density with an OAM $\ell_n = -1$, we have
\begin{eqnarray}
a_w(r,z,\varphi, t=0) &=& A_w \, e^{-r^2/w_0^2} \,\frac{1}{\sqrt{\pi}} \frac{\sqrt{2} r}{w_0} \, e^{i \varphi}, \label{aw2}\\
n(r,z,\varphi, t=0) &=& N e^{(z-z_0)^2/\sigma_z^2}\, e^{-r^2/w_0^2} \,\frac{1}{\sqrt{\pi}} \frac{\sqrt{2} r}{w_0} \, e^{-i \varphi}. \label{n2}
\end{eqnarray}
For these initial configurations, the spatial evolution of the three wave process during the same time of interaction as in the previous situation is given in detail in Fig.3. In this figure, we could observe that the electron density perturbation and optical wiggler have a similar longitudinal behavior as in the previous scenario with the maximum intensity of the perturbed electron density at the minimum of the wiggler amplitude. On the top of Fig.3(a), the transverse cross-section of the squared real part of the electron density perturbation, Re$\{|n|\}^2$, taken at the maximum value of $|n|$ is shown. As we can see, the orbital angular momentum of the beam is conserved while the beam undergoes a rearrangement in its transverse waist, leading to a growth in the vortice of the donut shape density perturbation represented in the lower transverse face of the box in Fig.3(a). From Fig.3(c), we can observe effectively the coherent stimulation and amplification of the scattered radiation,  which has the same longitudinal behavior as pointed out in the previous situation which now carrying an OAM with $|\ell_s| = 2$ according to the matching condition given by Eq.(\ref{matching}). This can be confirmed from the transverse top face of the box in Fig.3(c), where the squared real part of the scattered radiation, Re$\{a_s\}^2$, has four symmetrically well separated maxima.
When we compare the initial configurations given by Eqs.(\ref{aw1})-(\ref{n1}) and by Eqs.(\ref{aw2})-(\ref{n2}), we can observe that, in the former case, the leading pulse of the scattered radiation gets a greater amplification and compression than the last one. These behaviors are in part due to the overlapping factor $R$, which  is greater in the first configuration than the second one. This feature is also discussed in two alternative FEL descriptions as pointed out in Refs.\cite{hemsing2008, sprangle2009} with higher OAM modes \\

\section{Conclusions}

Using the quantum fluid model for a FEL, a set of paraxial nonlinear coupled equations, which describes the FEL instability as a three-wave interaction process, was derived and solved numerically for two initial configurations. In the first scenario, a constant Gaussian optical wiggler with a null OAM ($\ell_w = 0$) and a seed mode (scattered radiation) with OAM ($\ell_s = 1$) and a Gaussian profile in the propagation direction $z$. As pointed out in the one dimensional model described in Ref. \cite{guarumo}, it is observed the amplification and compression of the leading pulse of the scattered radiation accompanied by a growth of the electron density perturbation, while the optical wiggler is depleted. In addition, the leading pulse is transversely compressed, which opens the possibility of reaching high intensity output radiation with OAM, as obtained in a backward Raman amplifier\cite{titonature}. We remark that, as in the backward Raman amplifier, the increase of the leading pulse intensity is subjected to several kinds of nonlinearities, such as the filamentation effect in the FEL dynamics, as explained in Ref.\cite{rizzato} for an optical wiggler FEL. According to the OAM  matching condition given by Eq.(\ref{matching}), it is possible to show that electron density perturbation can acquire an OAM, opening a new possibility to design a helical electron beam.
In the second alternative scenario, we consider a constant optical wiggler with an OAM ($\ell_w = 1$) interacting with an electron density perturbation with an OAM ($\ell_n = -1$). In this case, the stimulated output radiation is generated during the interaction with amplification and compression. This scattered radiation during the interaction gets an OAM with $|\ell_s| = 2$ according to the   matching condition derived in Eq.(\ref{matching}). Also, it could be noted that the overlapping factor $R$ plays a crucial role in determining the coupling efficiency between the modes to obtain an intense coherent output radiation.
The numerical data visualization were done using the software VAPOR17\cite{vapor}.

\begin{acknowledgments}
We would like to thank the financial support of CAPES Brazil, and of the European Programme IRSES.
\end{acknowledgments}

\end{document}